\documentclass[
 aps, pra,
 amsmath,amssymb,
 11pt,
 final,
tightenlines,
 twoside,
 onecolumn,
 nofloats,
nofootinbib,
 superscriptaddress,
showkeys,
showkeywords,
 ]
{revtex4-2}

\usepackage[T2A]{fontenc}
\usepackage[utf8x]{inputenc}
\usepackage[russian,english]{babel}
\usepackage{graphicx}
\usepackage{dcolumn}
\usepackage{bm}
\input{maik.rty}

\setcitestyle{authoryear,round}
\def\squareforqed{\hbox{\rlap{$\sqcap$}$\sqcup$}}

\def\sq{\ifmmode\squareforqed\else{\unskip\nobreak\hfil
\penalty50\hskip1em\null\nobreak\hfil\squareforqed
\parfillskip=0pt\finalhyphendemerits=0\endgraf}\fi}

\def\utw{\smash{\rlap{\lower5pt\hbox{$\sim$}}}}

\def\udtw{\smash{\rlap{\lower6pt\hbox{$\approx$}}}}

\def\diameter{{\ifmmode\mathchoice
{\ooalign{\hfil\hbox{$\displaystyle/$}\hfil\crcr
{\hbox{$\displaystyle\mathchar"20D$}}}}
{\ooalign{\hfil\hbox{$\textstyle/$}\hfil\crcr
{\hbox{$\textstyle\mathchar"20D$}}}}
{\ooalign{\hfil\hbox{$\scriptstyle/$}\hfil\crcr
{\hbox{$\scriptstyle\mathchar"20D$}}}}
{\ooalign{\hfil\hbox{$\scriptscriptstyle/$}\hfil\crcr
{\hbox{$\scriptscriptstyle\mathchar"20D$}}}}
\else{\ooalign{\hfil/\hfil\crcr\mathhexbox20D}}%
\fi}}

\newcommand{\HII}{H{\sc ii}}

\begin{document}

\selectlanguage{english}




\title{ANALYSIS OF THE INTERSTELLAR MATTER AT THE PERIPHERY OF THE SUPERSHELL SURROUNDING THE CYG OB1 ASSOCIATION IN 2.12 MICRON MOLECULAR HYDROGEN LINE}

\author{\firstname{D.~S.}~\surname{Wiebe}}
\email{dwiebe@inasan.ru}
\affiliation{Institute of Astronomy (INASAN), Russian Academy of Sciences, Pyatnitskaya str. 48, Moscow 119017, Russia}

\author{\firstname{T.~G.}~\surname{Sitnik}}
\affiliation{Lomonosov Moscow State University, Sternberg Astronomical Institute, Universitetsky pr. 13, Moscow 119992, Russia}

\author{\firstname{A.~S.}~\surname{Rastorguev}}
\affiliation{Lomonosov Moscow State University, Sternberg Astronomical Institute, Universitetsky pr. 13, Moscow 119992, Russia}
\affiliation{Lomonosov Moscow State University, Faculty of Physics, Leninskie Gory 1 b.2, Moscow 119992, Russia}

\author{\firstname{T.~A.}~\surname{Lozinskaya}}
\affiliation{Lomonosov Moscow State University, Sternberg Astronomical Institute, Universitetsky pr. 13, Moscow 119992, Russia}

\author{\firstname{A.~M.}~\surname{Tatarnikov}}
\affiliation{Lomonosov Moscow State University, Sternberg Astronomical Institute, Universitetsky pr. 13, Moscow 119992, Russia}
\affiliation{Lomonosov Moscow State University, Faculty of Physics, Leninskie Gory 1 b.2, Moscow 119992, Russia}

\author{\firstname{A.~A.}~\surname{Tatarnikova}}
\affiliation{Lomonosov Moscow State University, Sternberg Astronomical Institute, Universitetsky pr. 13, Moscow 119992, Russia}

\author{\firstname{A.~P.}~\surname{Topchieva}}
\affiliation{Institute of Astronomy (INASAN), Russian Academy of Sciences, Pyatnitskaya str. 48, Moscow 119017, Russia}

\author{\firstname{M.~V.}~\surname{Zabolotskikh}}
\affiliation{Lomonosov Moscow State University, Sternberg Astronomical Institute, Universitetsky pr. 13, Moscow 119992, Russia}

\author{\firstname{A.~A.}~\surname{Fedoteva}}
\affiliation{Lomonosov Moscow State University, Sternberg Astronomical Institute, Universitetsky pr. 13, Moscow 119992, Russia}

\author{\firstname{A.~A.}~\surname{Tatarnikov}}
\affiliation{Lomonosov Moscow State University, Sternberg Astronomical Institute, Universitetsky pr. 13, Moscow 119992, Russia}
\affiliation{Lomonosov Moscow State University, Faculty of Physics, Leninskie Gory 1 b.2, Moscow 119992, Russia}

\begin{abstract}
We present observations of the vdB~130 cluster vicinity in a narrow-band filter centered at a $2.12\,\mu$m molecular hydrogen line performed at the Caucasus Mountain Observatory of the Lomonosov Moscow State University. The observations reveal an H$_2$ emission shell around vdB~130, coincident with a bright infrared shell, visible in all \textit{Spitzer} bands. Also, numerous H$_{2}$ emission features are detected around infrared Blobs E and W and in the vicinity of a protocluster located to the east of the shell, in a tail of a cometary molecular cloud. H$_2$ emission in the vicinity of the vdB~130 cluster is mostly generated in well-developed \HII\ regions and is of fluorescent nature. In the protocluster area, isolated spots are observed, where H$_2$ emission is collisionally excited and is probably related to shocks in protostellar outflows. Obtained results are discussed in the context of possible sequential star formation in the vicinity of the vdB~130 cluster, triggered by the interaction of the expanding supershell surrounding the Cyg OB1 association with the molecular cloud and an associated molecular filament.
\end{abstract}

\maketitle

\selectlanguage{english}

\section{Introduction}

\defcitealias{sit15}{Paper~I}
\defcitealias{sit19}{Paper~II}
\defcitealias{sit20}{Paper~III}

\begin{figure}
\centering{\includegraphics[width=0.9\textwidth]{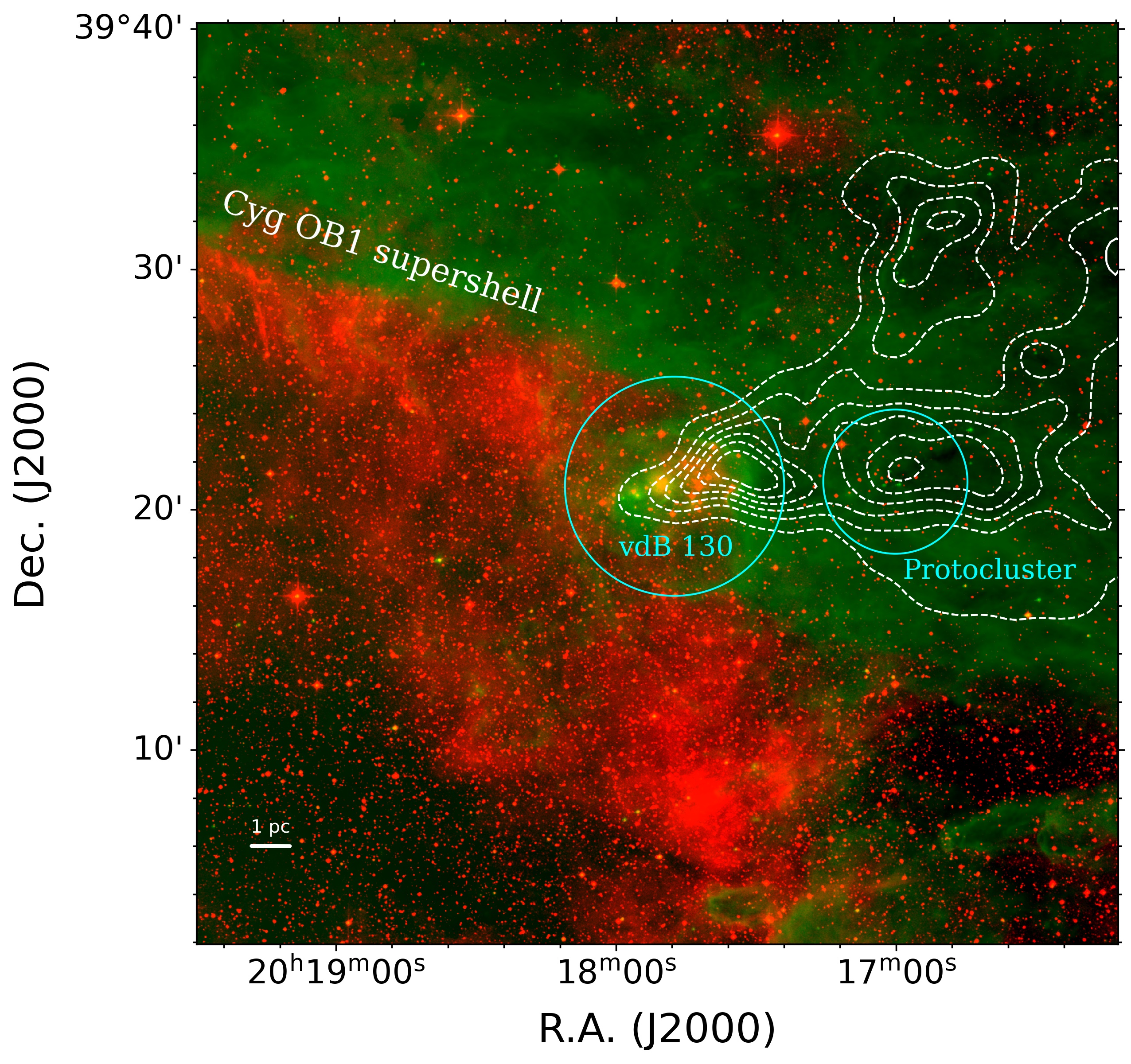}}
\caption {Northwestern part of the supershell surrounding the stellar association Cyg~OB1. Superimposed DSS2 red (red) and \textit{Spitzer} 8~$\mu$m (green) maps are overlaid with the integrated $^{13}$CO (1--0) emission contours outlining the molecular cloud \citep{sch07}. Contours run from 1 K $\cdot$ km s$^{-1}$ with a step of 2 K $\cdot$ km s$^{-1}$. The circles mark the central part of the vdB~130 cluster and the protocluster region. A small bar in the lower left corner corresponds to 1 pc at a distance of 1.67 kpc.}
\label{fig:cloud-proto}
\end{figure}

The study of the massive star formation complex in Cygnus has a long history and goes back to \citet{Blaha&Humphreys_1989}, who divided the sample of OB-stars into five large groups based on their positions on the sky. One of these groups is the well-known Cyg OB1 stellar association, which contains at least 50 OB stars and is surrounded by a large expanding supershell, having a size of $3^{\circ}\times4^{\circ}$ \citep{Humphreys}. This work concludes a series of our studies (\citealt{sit15,sit19,sit20}, hereinafter referred to as Paper I, II, and III, and \citealt{tatar16}) of the star formation in a region ($\alpha =20^{\rm h}16^{\rm m} - 20^{\rm h}18^{\rm m}$, $\delta=39^{\circ}15'-39^{\circ}35'$), which is located in the vicinity of the young embedded cluster vdB~130, next to a wall of the supershell{ The overall outline of the region is shown in Fig.~\ref{fig:cloud-proto}. Superimposed Digitized Sky Survey (DSS2)\footnote{\url{https://archive.stsci.edu/cgi-bin/dss_form}} red and \textit{Spitzer}\footnote{\url{https://sha.ipac.caltech.edu/applications/Spitzer/SHA/}} 8~$\mu$m maps are overlaid with the integrated $^{13}$CO (1--0) emission contours outlining the molecular cloud \citep{sch07}.}

A cometary molecular cloud, which has been designated as Cloud~A in~\citet{sch07}, stands out in the considered part of the supershell (Fig.~\ref{fig:cloud-proto}). Using expressions from \citet{1998A&A...335.1049S}, we estimated its mass to be $\sim3000\,M_{\odot}$ within a 1 K $\cdot$ km s$^{-1}$ level (observational data have been kindly provided by N. Schneider). Also, \citet{sch07} have found a velocity gradient of 0.5 km s$^{-1}$ pc$^{-1}$ along a major axis of the cloud and suggested that Cloud~A is most likely shaped by UV-radiation and stellar winds from nearby OB stars. The cloud is indeed extended toward an illuminating source, that is, in the direction of the closest OB stars in Cyg OB1. The size of the molecular cloud is $0.4^{\circ}$ or 12~pc at the adopted distance of $\sim 1.67 \pm 0.06$ kpc toward the cluster and the association \citep{rastorguev_companion_paper}.

 The cloud is associated with two star-forming sites. A first one, young vdB~130 cluster, is located in the head of the cloud (a pillar), and another one, a compact protocluster, is located in the cloud's tail (Fig.~\ref{fig:cloud-proto}, see also Fig.~11 in \citetalias{sit19}). A physical relation of the Cyg~OB1 association, supershell, the vdB~130 cluster, and the molecular cloud follows from estimated distances and radial velocities as well as from some indirect factors (\citetalias{sit15}). An age of vdB~130 does not exceed 10~Myr (\citetalias{sit15} and \citetalias{sit20}). Its youth is consistent with the presence of Class~I and Class~II protostars \citep{Kuhn+2021}, which are visible both in the vdB~130 area and in the protocluster area.

\begin{figure}
\centering{\includegraphics[width=0.8\textwidth]{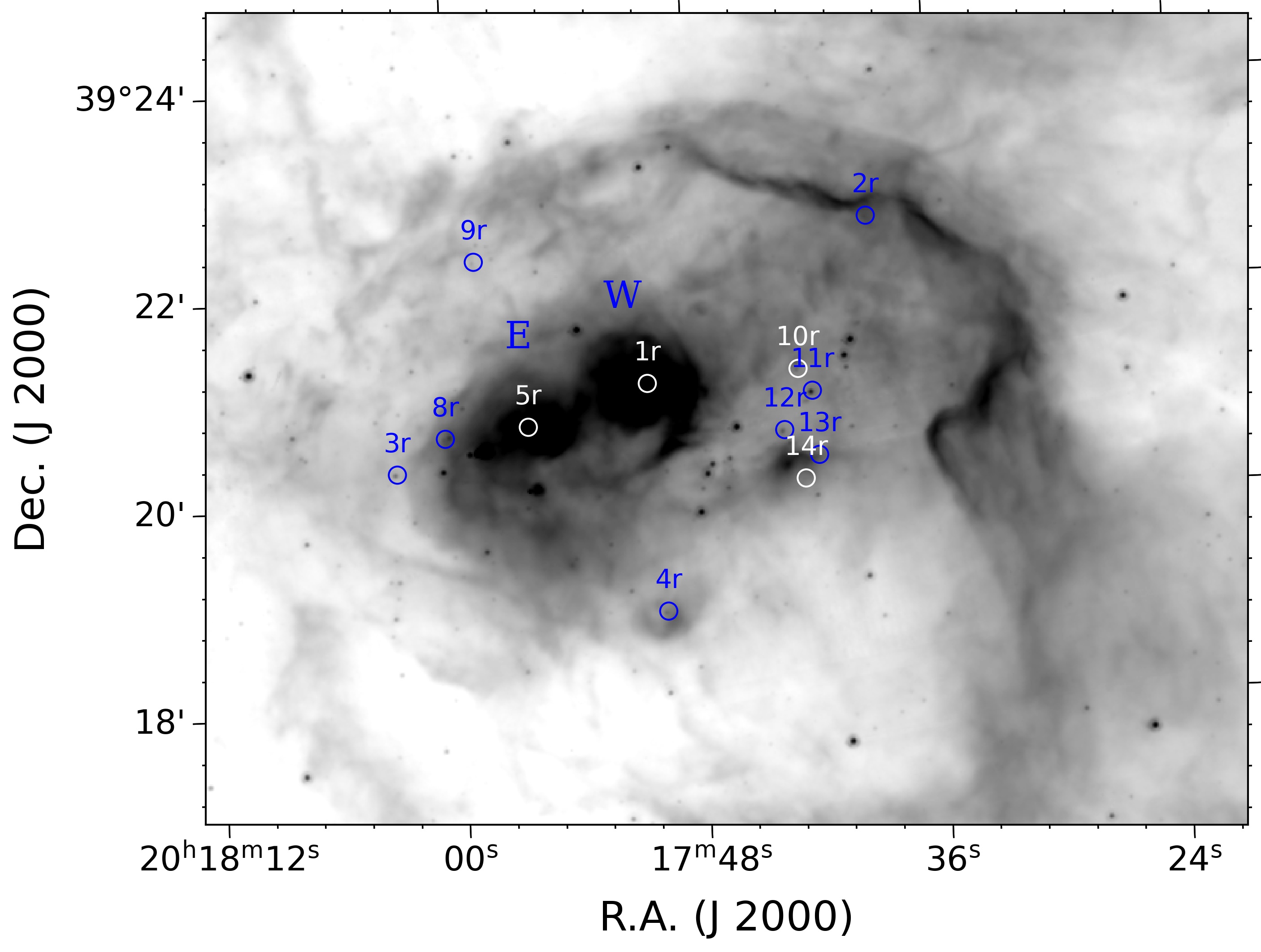}
\includegraphics[width=0.8\textwidth]{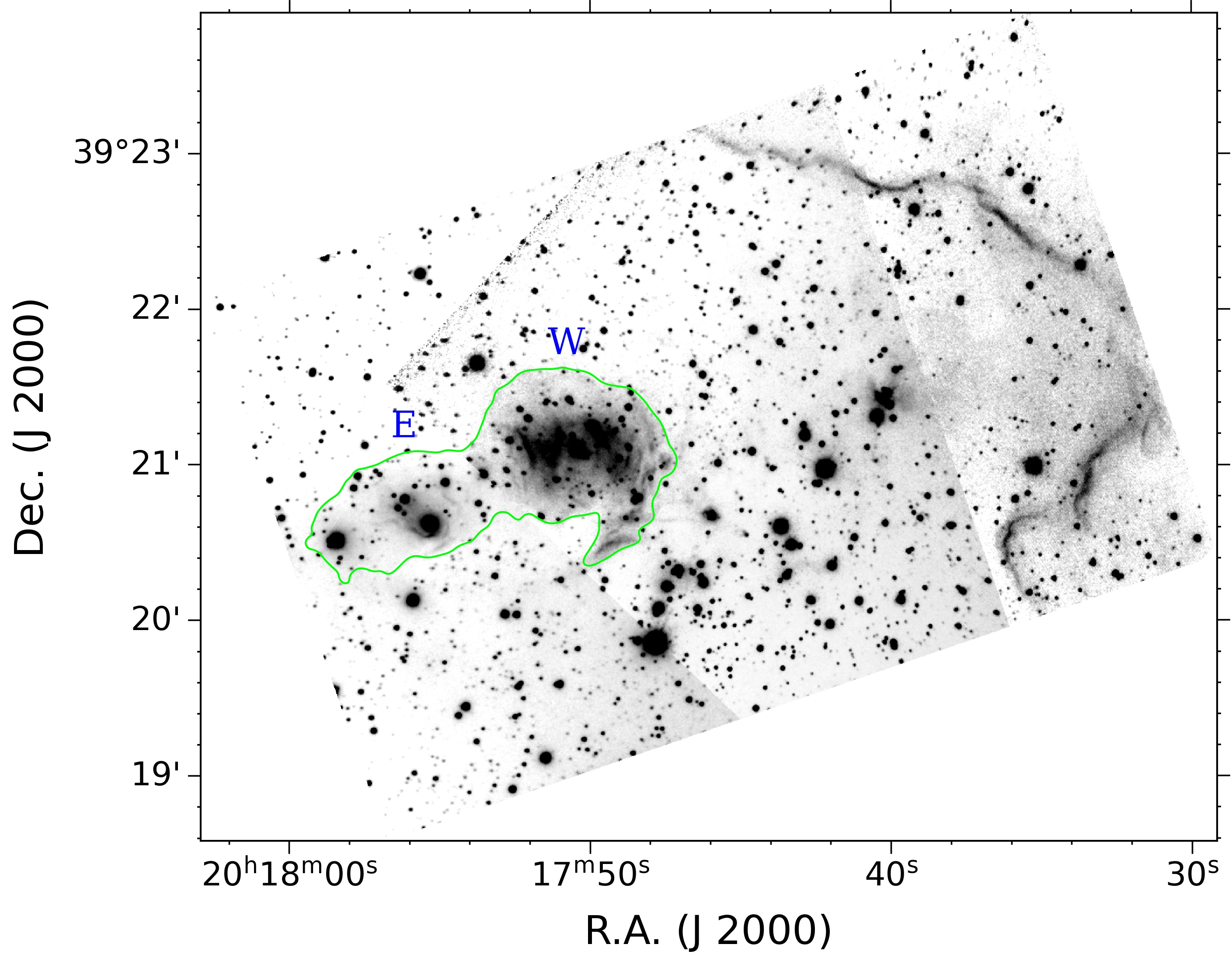}}
\caption{Images of the vdB~130 vicinity in the \textit{Spitzer} 8\,$\mu$m band (top) and in molecular hydrogen line ($\lambda =2.12\,\mu$m) (bottom). Infrared Blobs E and W along with several stars from the original \citet{rac74} list are indicated on the top panel. Blobs' approximate contours are also shown in the bottom panel.} \label{fig:vdB130}
\end{figure}

Analysis of interstellar extinction toward vdB~130 has shown that foreground material is characterized by the normal extinction law with $R_{v} = 3.1$ (\citealt{tatar16}, \citetalias{sit20}). Absorption within the cluster itself is very large and inhomogeneous. $R_{v}$ toward central stars in Blob E and Blob W (Fig.~\ref{fig:vdB130}, top) can be as high as 8 according to our data and data from \citet{rac74}. From the west, the vdB~130 cluster is surrounded by an IR shell, having a size of 3~pc and discernible in all the IRAC bands and in $24\,\mu$m MIPS band of the {\it Spitzer} telescope (Fig.~\ref{fig:vdB130}, top).

One of the major large-scale motions of the interstellar material in the studied region is the expansion of the supershell under the influence of wind and ionizing radiation of Cyg OB1 association stars. The expanding supershell compresses pre-existing molecular clumps or clumps emerging in the swept-up material \citep[e.g.,][]{Deharveng}. The expansion speed of the supershell, derived from the analysis of interferometric observations in the H$\alpha$ line, reaches 80~km/s in some directions \citep{loz98,arkhipova}. 

It is tempting to suggest that a cometary shape of the molecular cloud (with its head inside the supershell and its tail outside of it) as well as two star-forming regions (the vdB~130 cluster and the protocluster), embedded in the head and tail of the cloud, are all signatures of sequential star formation triggered by the supershell expansion (Fig.~\ref{fig:cloud-proto}, see also \citetalias{sit19} and \citetalias{sit20}). This region is convenient for identifying the consequences of such an expansion, since it is located at the periphery of the supershell, and the suggested process of triggered star formation ``unfolds'' in the plane of the sky.

Some arguments in favor of the sequential star formation in this region have been given in our previous studies. Here we present results of our observations of the vicinity of the vdB~130 cluster and the protocluster in the infrared (IR) filter centered at 2.12-$\mu$m molecular hydrogen line, installed at the 2.5-m telescope of the Caucasus Mountain Observatory of the Sternberg Astronomical Institute, Lomonosov Moscow State University (CMO SAI MSU).

This line is used quite often as a tool to probe physical conditions in protostellar outflows and photodissociation regions (PDRs). Being among the brightest H$_2$ ro-vibrational lines, it can be excited in two competing ways. Collisional excitation prevails in shocks, while radiative excitation is more typical for PDRs, where material is irradiated by nearby massive stars. Due to the lack of the line profile information, in this study we only consider H$_2$ emission morphology and its relation to other tracers.

The structure of the paper is the following. In Section~2 the equipment and methods for processing observations in the H$_{2}$ line as well as the used archived data are described. Section~3 presents an analysis of new observational data in the vicinity of IR shell and protocluster. Possible manifestations of sequential star formation, triggered by the expanding supershell, are discussed in Section~4. In Conclusion the results of the study are summarized.

\section{Observations and data reduction}

This study is based on IR observations performed with the ASTRONIRCAM infrared camera \citep{Nadjip17} installed on the \mbox{2.5-m} telescope (CMO SAI MSU), from April 3rd to September 1st, 2020. FWHMs of star images varied from $0.8''$ to $1.2''$. The ASTRONIRCAM frame size is $1024\times1024$ pixel with an angular scale of $0.27''$/pixel and a field of view of $280''\times280''$. We used the H$_2$ v=1--0 (S1) filter with $\lambda_0=2.132\,\mu$m and FWHM $=0.046\,\mu$m and the Kcont filter with $\lambda_0=2.273\,\mu$m and FWHM $=0.039\,\mu$m. The first one is centered on v=1--0 (S1) molecular hydrogen line and the second one allows measuring the continuum level in the neighboring spectrum region. Each final frame represents a sum of many separate frames, obtained with short exposures ($\sim9$~s). Between these frames, the telescope was shifted by an angle of about $3-5''$. A total exposure times for final frames is from 910 to 3940~s. A sky background has been measured separately at the same air mass. Surface brightnesses and standard deviations for final frames were estimated using 2MASS stars as flux calibrators \citep{Tatarnikov2023}. Standard deviations are $1.5\cdot10^{-17}$ erg/s/cm$^2$/pixel for the shell area, $1.3\cdot10^{-17}$ erg/s/cm$^2$/pixel for the blobs' area, and $8.8\cdot10^{-18}$ erg/s/cm$^2$/pixel for the protocluster area. These numbers can be considered as upper estimates for the sensitivity limit of our data. In the following we present data for the H$_2$ filter with the Kcont filter data subtracted under the simplified assumption that the continuum is flat in the region of the spectrum around 2\,$\mu$m.

Interstellar absorption is not taken into account. It can be important at these wavelengths \citep{2022arXiv220608245H}, but given the complicated structure of extinction in the area, its taking into account is non-trivial. We believe that its neglect would not change our conclusions, as we are mostly interested in the emission morphology. All the considered data have been regridded to the same astrometry, corresponding to H$_2$ data. In some cases, described below, H$_2$ images have been convolved to the resolution of the {\it Spitzer} $8\,\mu$m data (FWHM $2''$, \citealt{2004ApJS..154...10F}) using a Gaussian kernel provided by \citet{2011PASP..123.1218A}.

We also use our H$\alpha$ observations of the region presented in \citetalias{sit15} and archival data obtained with \textit{Spitzer}.

\section{Molecular hydrogen in the vicinity of the vdB~130 cluster and the protocluster}

\begin{figure}[b!]
\centering{\includegraphics[width=0.8\textwidth]{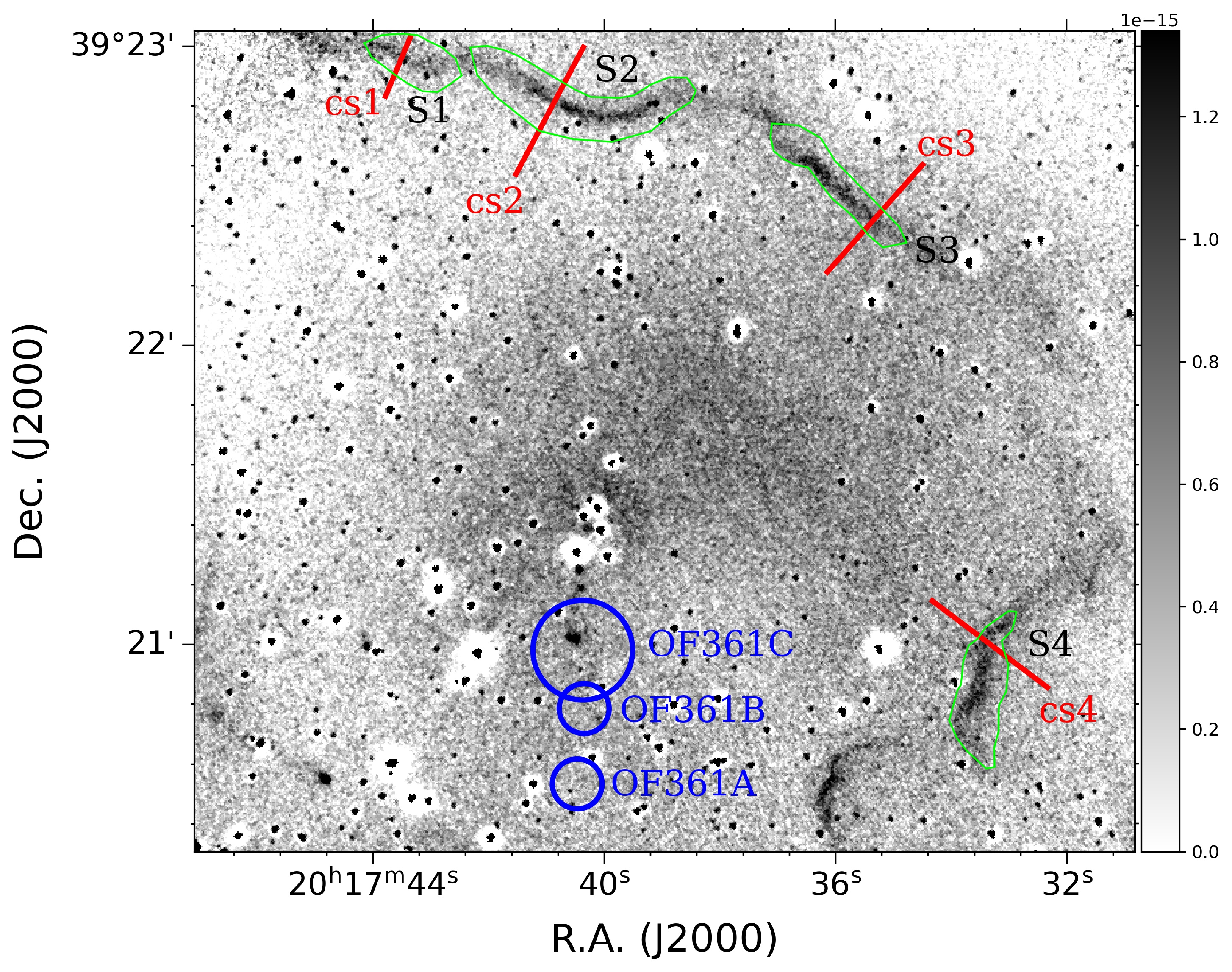}
\includegraphics[width=0.8\textwidth]{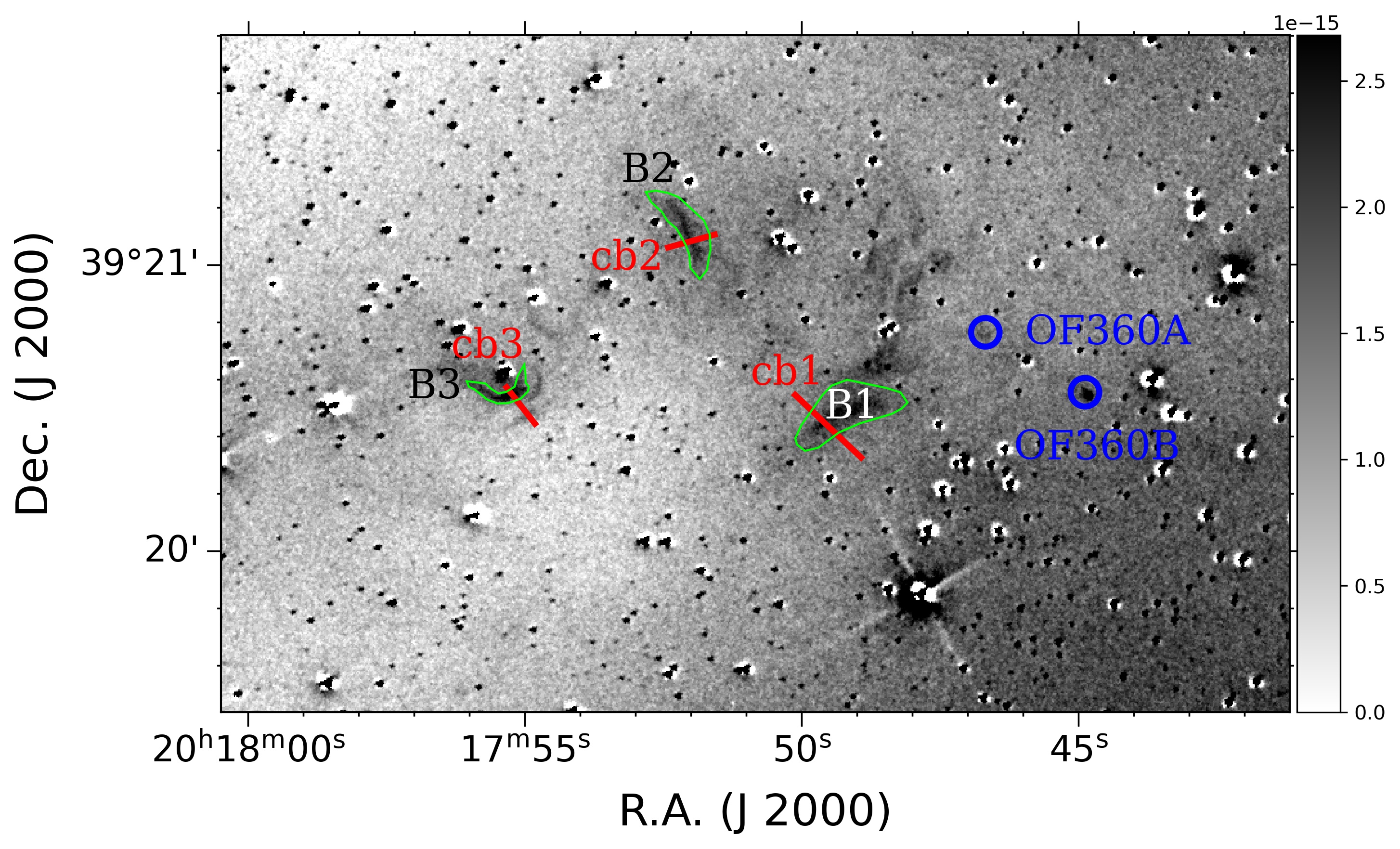}}
\caption{Molecular hydrogen emission in the vicinity of vdB~130 (erg/s/cm$^2$/arcsec$^2$, linear scale, Kcont continuum is subtracted). Top: IR shell area. Bottom: blobs' area. Green contours and red lines indicate regions and cuts selected for detailed analysis. Also indicated are outflows OF360 and OF361, identified previously by \citet{2018ApJS..234....8M}.}
\label{fig:H2emisshellblobs}
\end{figure}

Molecular hydrogen emission is clearly visible in several locations across the studied region (Fig.~\ref{fig:vdB130}, bottom). It neatly traces the infrared shell around the vdB~130 cluster, surrounds infrared blobs, specifically Blob~W and Blob~E, and appears as several compact spots in the protocluster region. In the top panel of Fig.~\ref{fig:H2emisshellblobs} we show the molecular hydrogen emission map for the IR shell vicinity with Kcont continuum subtracted, highlighting four bright H$_2$ emission features, denoted S1, S2, S3, and S4, along with four cross-cuts (cs1, cs2, cs3, cs4), which will be discussed below. The bottom panel of Fig.~\ref{fig:H2emisshellblobs} shows features (B1, B2, B3) and cuts (cb1, cb2, cb3) selected for the analysis in the blobs' area.

\begin{figure}
\centering{\includegraphics[width=0.8\textwidth,clip=]{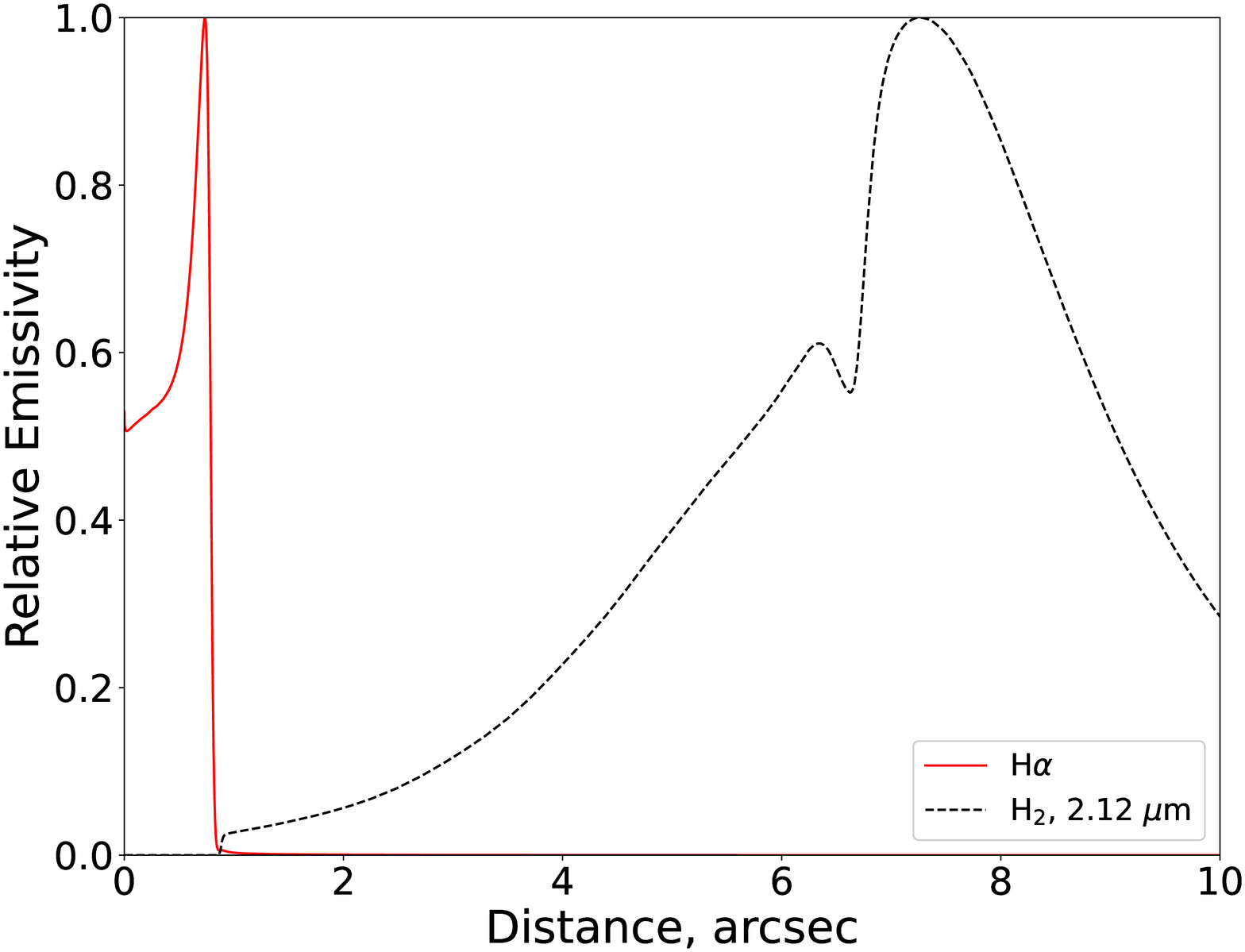}}
\caption{Theoretical H$\alpha$ and H$_2$ intensity profiles across the shell of an ionized hydrogen region. The region is ionized by an O-type star located at 1 pc from the inner border of the computational domain (model 4O1). The gas density in the shell is $10^4$~cm$^{-3}$.}
\label{theordiag}
\end{figure}

In order to relate observed quantities to each other and to suggested morphological features of the region, we use the Cloudy code \citep{2017RMxAA..53..385F}{, which allows computing a structure of an HII region and an adjacent photodissociation region with a very detailed account of most relevant processes. Specifically, we calculated a number of models for a shell with a central star having an effective temperature of 25000\,K (B-star, $\log g=3.9$, $\log L/L_\odot=3.7$) or 50000\,K (O-star, $\log g=4.1$, $\log L/L_\odot=6.1$). Corresponding stellar models have been downloaded from the TLUSTY Web-site\footnote{\url{http://tlusty.oca.eu/Tlusty2002/tlusty-frames-cloudy.html}} \citep{2007ApJS..169...83L}. The star is located at a distance $d$ from the inner border of the modeled region, which is assumed to be uniform. Considered density values are $10^3$ and $10^4$\,cm$^{-3}$.

Results are presented in Table~\ref{cloudyresshell}. A first number in a model designation (a first column of Table~\ref{cloudyresshell}) is the density exponent, then comes the letter that designates the ionizing star spectral type, and the last number is the distance $d$ from the to the inner edge of the computational domain in parsecs. So, for example, a model 3B0.1 corresponds to a region with a gas number density of $10^3$\,cm$^{-3}$, and an inner edge located at a distance of 0.1~pc from a B star. An example of the theoretical shell structure for a model 4O1 is shown in Fig.~\ref{theordiag}. In this case, an edge of the \HII\ region is located at a distance of about $0.7''$ from an inner border of the computational domain. As the geometry of the gas and dust distribution in the vdB~130 vicinity} is far from being spherical, we do not try to reproduce emergent spectra, relying instead upon emissivities as functions of a distance from an ionising source. This limits available options to considering mutual locations of various ISM components and correlations between their respective intensities. Stars have been masked both in the shell area and the blobs' area.

\subsection{Infrared Shell}

\begin{figure}
\centering{\includegraphics[width=0.8\textwidth,clip=]{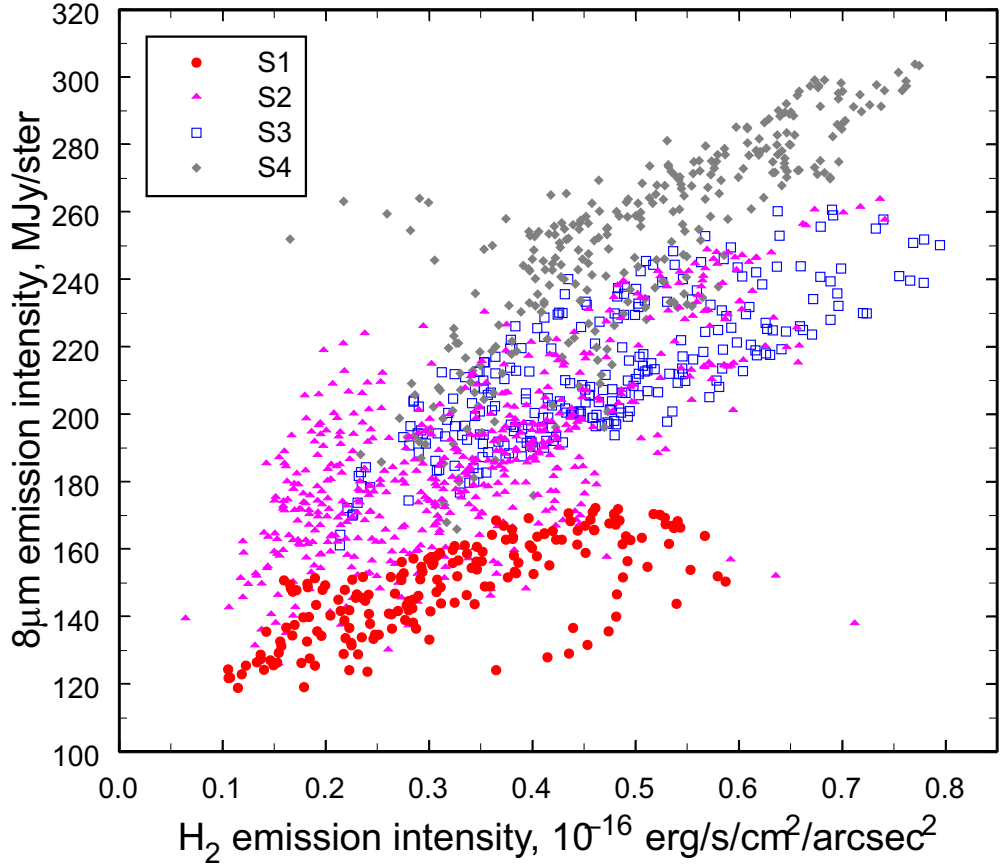}}
\caption{Relation between molecular hydrogen emission and 8\,$\mu$m emission in the selected regions of the IR shell.}
\label{fig:H2-8-shell}
\end{figure}

The H$_2$ emission from the shell itself is clearly discernible, along with diffuse foreground and/or background H$_2$ emission projecting onto the cluster area. The H$_2$ shell spatial location coincides with the location of 8\,$\mu$m filaments. Pixel-by-pixel comparison of H$_2$ emission and $8\,\mu$m emission for features S1--S4 is presented in Fig.~\ref{fig:H2-8-shell}. The H$_2$ data have been convolved to a resolution of the {\it Spitzer} $8\,\mu$m data. We see that the molecular hydrogen emission intensity correlates with 8\,$\mu$m emission intensity in all the four features, which hints that it is UV emission that excites both PAHs and hydrogen molecules.

\begin{figure}
\centering{\includegraphics[width=0.8\textwidth,clip=]{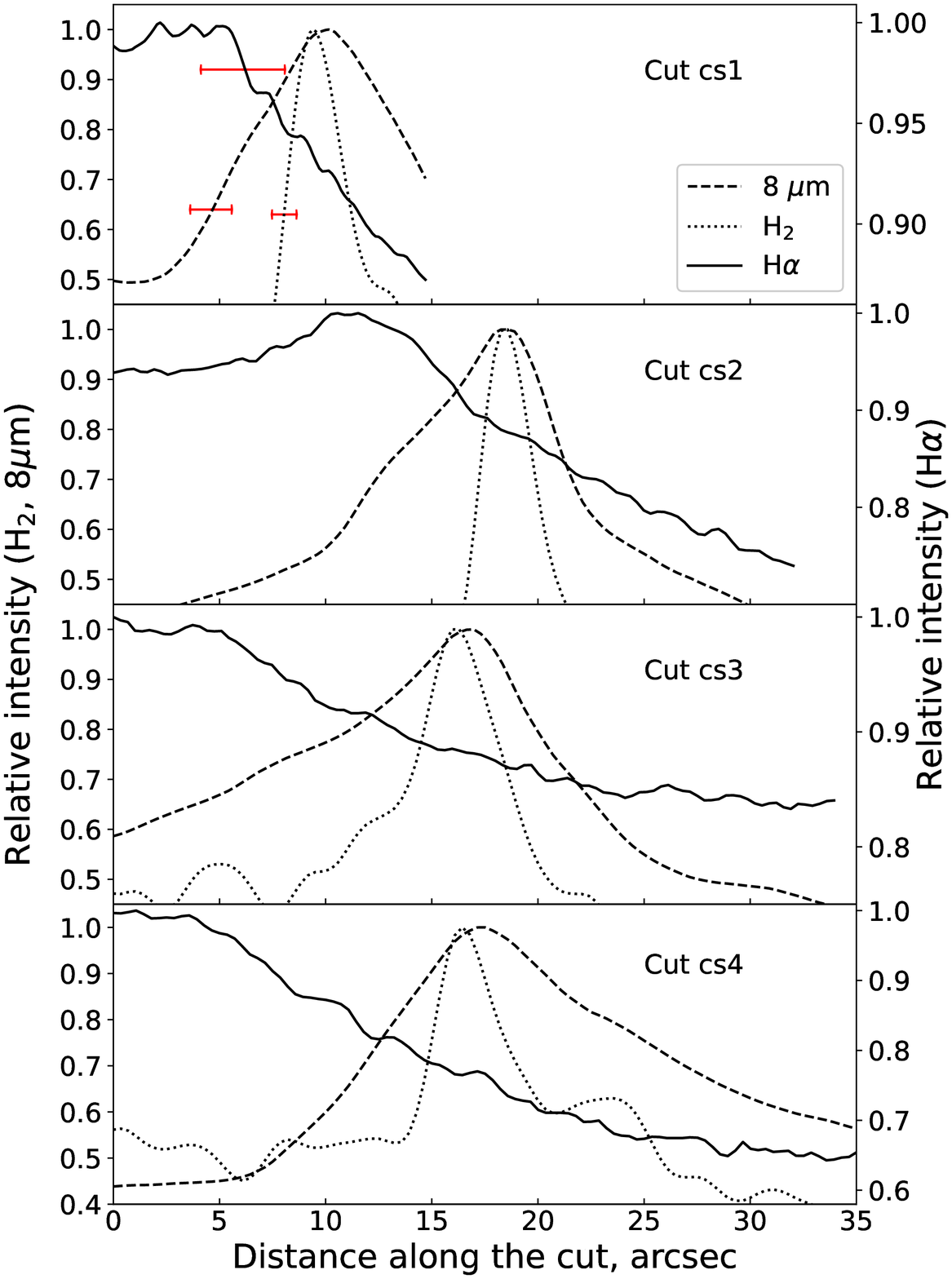}}
\caption{H$\alpha$, 8\,$\mu$m, and H$_2$ intensity profiles along the four cuts through the shell indicated in the top panel of Fig.~\ref{fig:H2emisshellblobs} (relative units).{ Red horizontal} bars on the top panel indicate respective angular resolutions, which are adopted to be $1''$ for H$_2$ data, $2''$ for $8\,\mu$m data, and $4''$ for H$\alpha$ data \citepalias{sit15}.}
\label{shellcuts}
\end{figure}

Observed intensity profiles along the shell cuts marked in the top panel of Fig.~\ref{fig:H2emisshellblobs} are shown in four panels of Fig.~\ref{shellcuts}. Here we again show H$_2$ data that have been convolved to a resolution of the {\it Spitzer} $8\,\mu$m data. It is seen that peaks of H$_2$ and PAH emission are nearly coincident and somewhat displaced relative to the border of the \HII\ region, which we, somewhat arbitrarily, identify with the location, where H$\alpha$ intensity starts decreasing. This displacement is of the order of $5''-10''$. The breadths of 8\,$\mu$m and H$_2$ emission shells are similar to each other and equal to $3''-5''$.

\begin{table}
\setlength{\tabcolsep}{1pt}
\caption{Modeled parameters of the PDR structure. The model designation consists of a density exponent, a letter indicating a star spectral type, and a distance (in parsec) from the star to the inner border of the modeled region.}
\label{cloudyresshell}
\begin{tabular}{l|r|c|c}
\hline
Model &$d$,& H$_2$ $2.12\,\mu$m--IF & H$_2$ shell   \\
      & pc&separation, $''$         & breadth, $''$ \\
\hline
3B0.1  & 0.1 & 35            & 30 \\
3B0.3  & 0.3 & 30            & 30 \\
\hline
4B0.1  & 0.1 & 6             &  4 \\
4B0.3  & 0.3 & 2             &  2 \\
\hline
3O0.1  & 0.1 & {\bf 30}            & {\bf 40} \\
3O0.3  & 0.3 & {\bf 40}            & 50 \\
3O1    & 1   & 60            & 50 \\
3O3    & 3   & 50            & {\bf 50} \\
3O10   & 10  & 10            & 30 \\
\hline
4O0.1  & 0.1 &  7            & 5  \\
4O0.3  & 0.3 &  7            & 5  \\
4O1    & 1   &  7            & {\bf 5}  \\
4O3    & 3   &  4            & 3  \\
\hline
\end{tabular}
\end{table}

These values can be related with the results of theoretical modeling, presented in Table~\ref{cloudyresshell}. We see that observed geometrical constraints in the shell are only satisfied in models with gas density of $10^4$\,cm$^{-3}$. At this density, O-type stars produce breadths of H$_2$ emission shells, which are similar to observed ones, as are the separations between the shells and the \HII\ region border, at all the considered distances. A B-type star located at a distance greater than $\sim0.3$ pc produces the shells that are too narrow and too close to the \HII\ region border.

Overall, we conclude that despite the fact that the IR shell apparently engulfs vdB~130, the cluster stars alone cannot be responsible for the shell structure, and it also ``feels'' the influence of the Cyg OB1 stars.

\subsection{Infrared Blobs}

In the bottom panel of Fig.~\ref{fig:H2emisshellblobs} we show H$_2$ emission distribution to the east of the IR shell, in the vicinity of Blob~E and Blob~W. Here the structure of the H$_2$ emission map is much more intricate. It consists of numerous curved filaments with their convex parts being oriented both toward and from the ionising stars 10r--14r (see \citetalias{sit20} and top panel of Fig.~\ref{fig:vdB130}).

\begin{figure}
\centering{\includegraphics[width=0.8\textwidth,clip=]{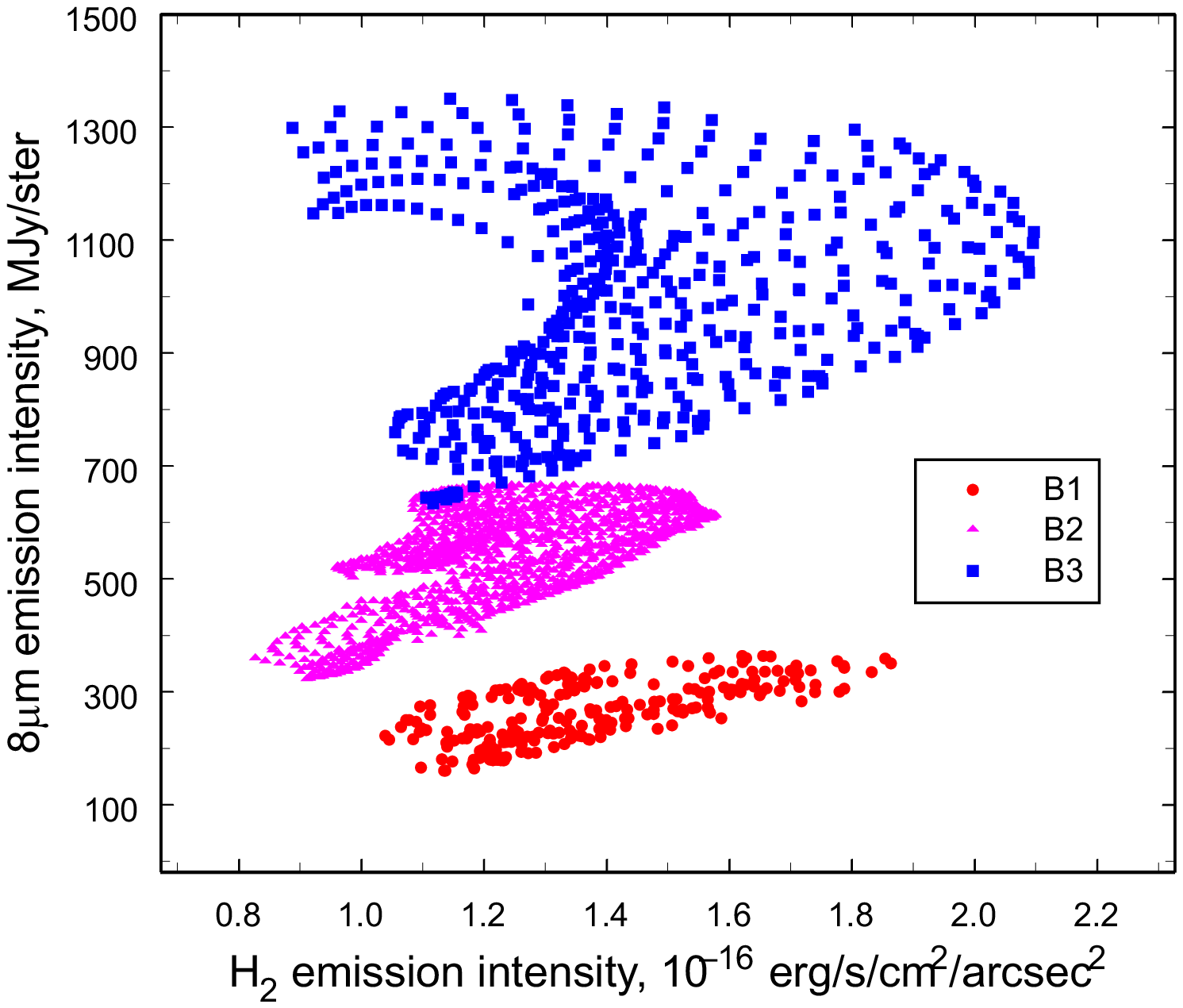}}
\caption{Relation between molecular hydrogen emission and 8\,$\mu$m emission in the selected regions of the blobs' area.}
\label{fig:H2-8-blobs}
\end{figure}

\begin{figure}
\centering{\includegraphics[width=0.8\textwidth,clip=]{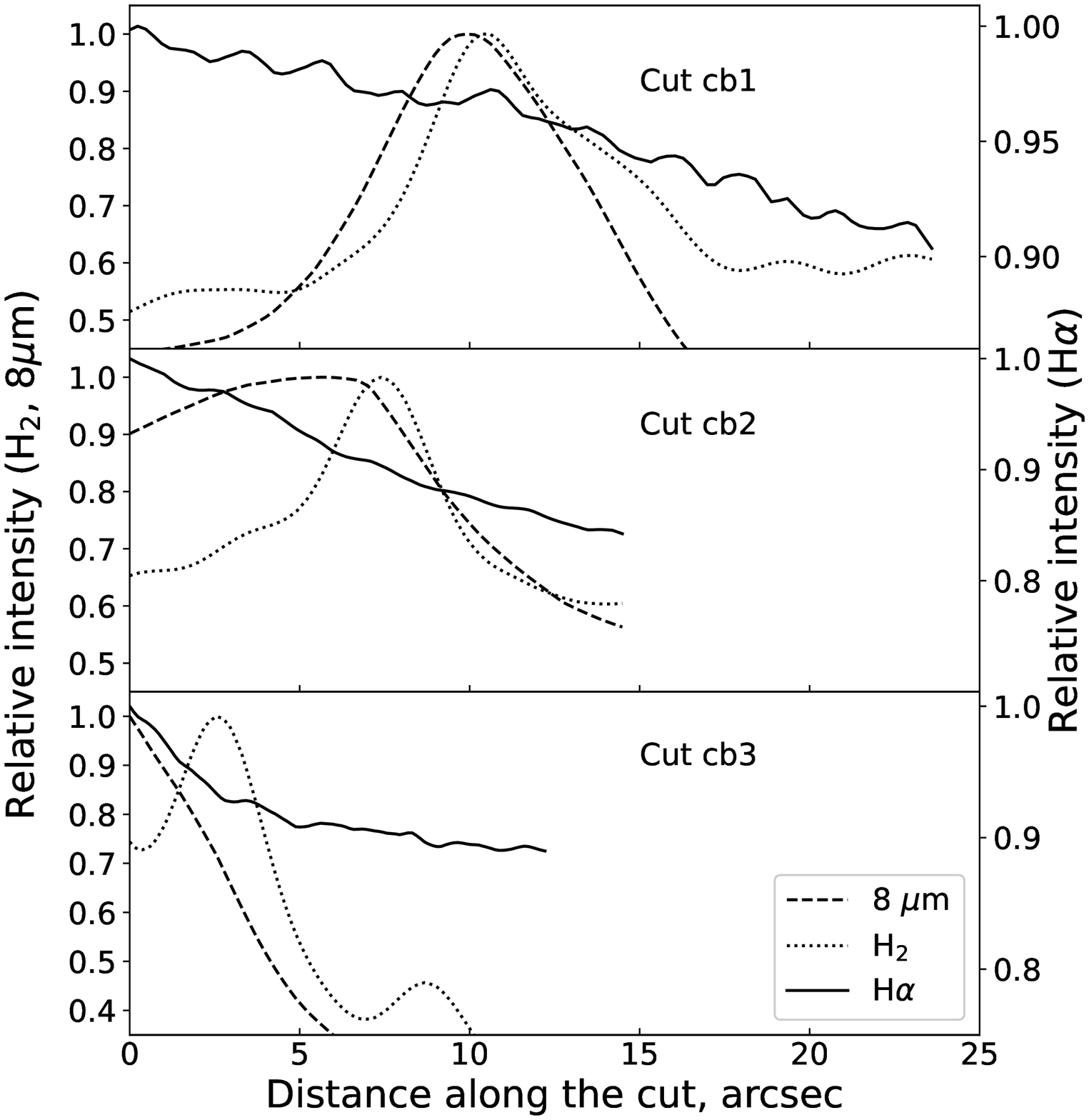}}
\caption{H$\alpha$, 8\,$\mu$m, and H$_2$ intensity profiles along the three cuts through the blobs indicated in the bottom panel of Fig.~\ref{fig:H2emisshellblobs} (relative units).}
\label{blobscuts}
\end{figure}

In Fig.~\ref{fig:H2-8-blobs} we demonstrate a relation between H$_2$ emission and 8\,$\mu$m emission for the features identified in Fig.~\ref{fig:H2emisshellblobs} (bottom panel). The H$_2$ data shown in Fig.~\ref{fig:H2-8-blobs} and Fig.~\ref{blobscuts} have been convolved to a resolution of the {\it Spitzer} $8\,\mu$m data.{ There is a positive correlation between the two tracers in the feature B1 (red circles), but in the brightest feature B3, the situation is more complicated, with the relation being clearly split into two branches (blue squares). A further analysis shows that the descending branch, with $8\,\mu$m emission decreasing as H$_2$ emission becomes brighter, corresponds to the inner part of the feature, while the ascending branch corresponds to its outer part. Such a structure is indicative of some displacement between the H$_2$ emission peak and the $8\,\mu$m emission peak. A hint to a similar branch splitting is also visible in the feature B2. This can also be a signature of a displacement between the H$_2$ and $8\,\mu$m emission peaks or be an indication of an even more complex morphology.}

In Fig.~\ref{blobscuts} we show profiles of various emission tracers along the cuts shown at the bottom panel of Fig.~\ref{fig:H2emisshellblobs}. Small sizes of the blobs and associated H$_2$ emission features along with insufficient resolution of available H$\alpha$ observational data only allow estimating upper limits to the separations between ionisation and dissociation fronts (as we cannot reliably determine the location of the ionisation front) and to the breadths of H$_2$ emission shells, which are of the order of few arcsec. In our modeling, this roughly corresponds to a B1V~star, located at a distance of about $0.1-0.3$ pc from the shell with a density of $10^3-10^4$ cm$^{-3}$, which is consistent with small sizes of the blobs and with our hypothesis that they represent \HII\ regions around the vdb~130 cluster stars 1r and 5r.

In Fig.~\ref{blobscuts} (bottom panel) we see a displacement between the $8\,\mu$m and H$_2$ emission maxima in the cb3 feature, mentioned above. A hint to a similar displacement is also seen in the cb2 feature (middle panel). It is tempting to assume that different spatial locations of $8\,\mu$m emission shell and H$_2$ emission shell indicate that H$_2$ molecules, unlike PAH molecules, are mostly collisionally excited in these features, so that we actually see a separation between an ionisation front and a shock front.

\subsection{Molecular hydrogen objects}

A region to the west of the IR shell (where the protocluster is located) does not host any extended H$_2$ emission features, but it harbors several compact fuzzy objects in our H$_2$ filter image. As there are no early-type stars in this area \citepalias{sit20}, the H$_2$ line emission requires collisional excitation, hinting at the presence of shock waves. Thus, these compact objects may correspond to molecular outflows formed by protostars, including massive ones \citep[see, for example,][]{wolf}. (Possibility of massive star formation in this region is supported by the presence of a feature reminiscent of an infrared dark cloud visible on a large-scale 8\,$\mu$m map.) And, indeed, the compact H$_2$ emission sources in the protocluster area have been marked as outflow features by \citet{2018ApJS..234....8M}.

\begin{figure}
\centering{\includegraphics[width=0.8\textwidth]{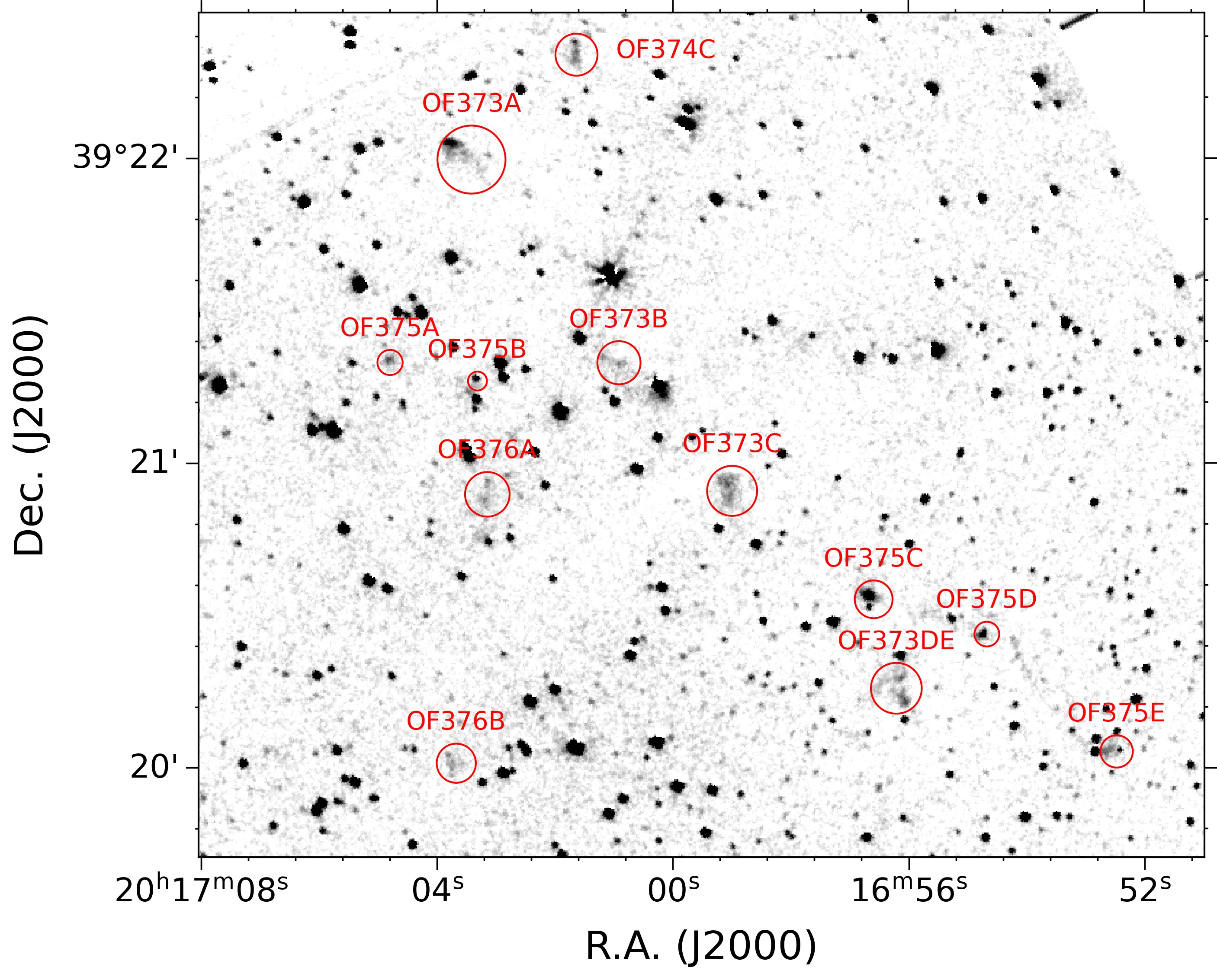}
\includegraphics[width=0.8\textwidth]{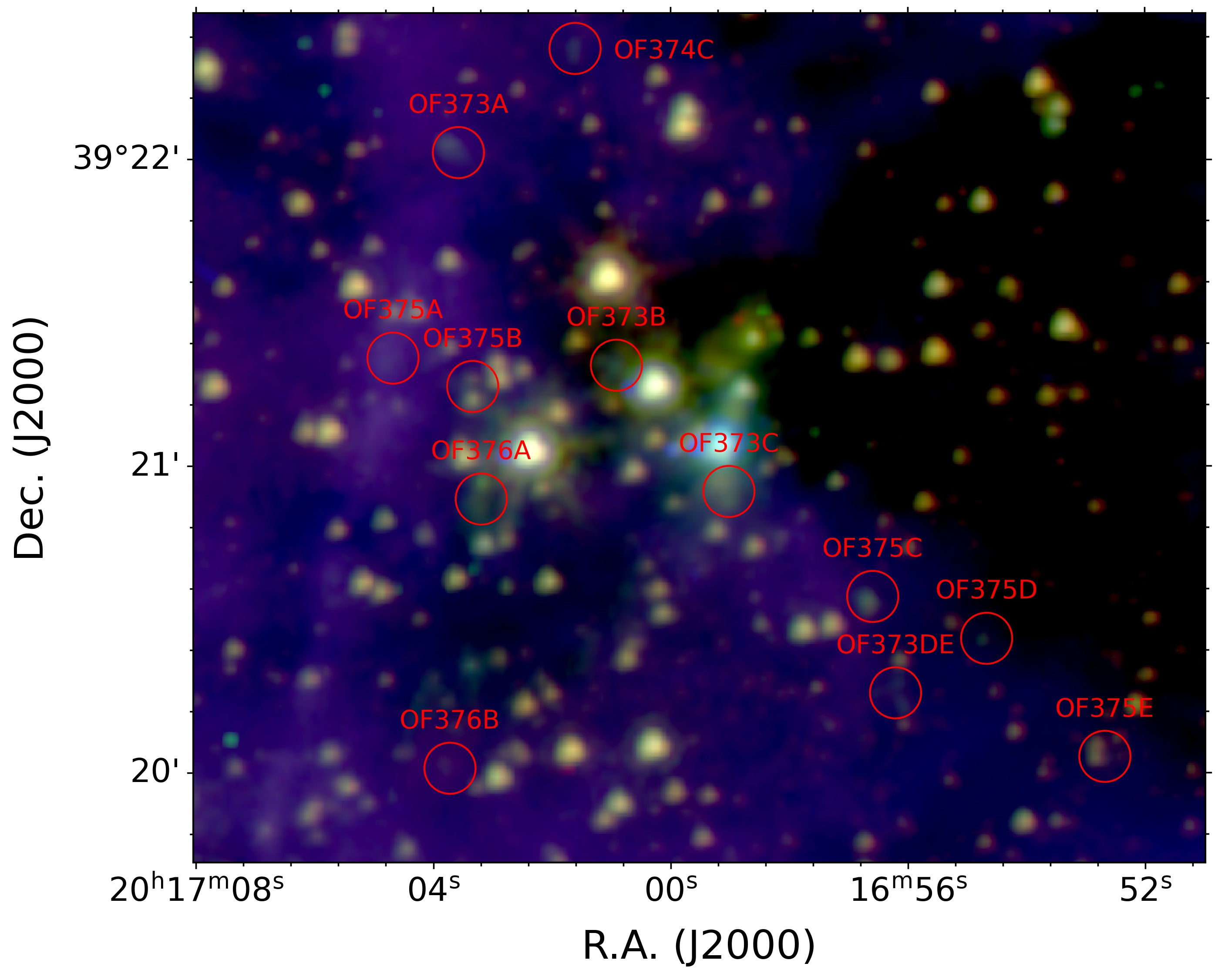}}
\caption{Top: a region of the protocluster core in 2.12 micron molecular hydrogen line after continuum subtraction. Shown are outflows features, identified by \citet{2018ApJS..234....8M}. Bottom: a color image of the same region composed from $3.6\,\mu$m (red), $4.5\,\mu$m (green), and $8\,\mu$m (blue) Spitzer bands.}
\label{fig:EGO}
\end{figure}

The entire Cygnus~X complex has been mapped in the molecular hydrogen 2.12\,$\mu$m line in the UWISH2 project \citep{2015MNRAS.454.2586F}. \citet{2018ApJS..234....8M} have used data from this survey to identify four potential outflows, OF373, OF374, OF375, and OF376 in their notation. These outflows are also associated with Molecular Hydrogen Emission-Line Objects (MHO) MHO~974, MHO~973, MHO~976, and MHO~977 from the MHO catalog \citep{2010A&A...511A..24D}\footnote{\url{http://astro.kent.ac.uk/~df/MHCat/index.html}}. The isolated features of these outflows, shown in Fig.~\ref{fig:EGO} (top), coincide with compact sources on our map. Coordinates and approximates sizes of the molecular hydrogen objects, identified by \citet{2018ApJS..234....8M} with outflow features and also observed in this work, are listed in Table~\ref{MHO_OF}.

\begin{table*}
\caption{Molecular hydrogen objects, identified by \citet{2018ApJS..234....8M} with outflow features and also observed in this work. In parentheses we indicate a panel number in Fig.~11 from \citet{2018ApJS..234....8M} with a corresponding map.}
\label{MHO_OF}
\begin{tabular}{l|c|c|c|c}
\hline                           
MHO ID & OF ID & $\alpha$ & $\delta$ & Size \\
\hline                           
MHO~973 (357)&OF374C  & 20:17:01.6 & +39:22:21  & $4''$\\
\hline
MHO~974 (356)&OF373A  & 20:17:03.6 & +39:22:01 & $7''$\\
       &OF373B  & 20:17:00.9 & +39:21:19  & $4''$\\
       &OF373C  & 20:16:59.0 & +39:20:55  & $5''$\\
       &OF373DE & 20:16:56.2 & +39:20:15  & $5''$\\
\hline                           
MHO~976 (358)&OF375A  & 20:17:04.7 & +39:21:21  & $2''$\\
       &OF375B  & 20:17:03.4 & +39:21:15  & $2''$\\
       &OF375C  & 20:16:56.6 & +39:20:34  & $4''$\\
       &OF375D  & 20:16:54.7 & +39:20:26  & $2''$\\
       &OF375E  & 20:16:52.7 & +39:20:03  & $3''$\\
\hline                           
MHO~977 (359)&OF376A  & 20:17:03.2 & +39:20:53  & $4''$\\
       &OF376B  & 20:17:03.7 & +39:20:01  & $4''$\\
\hline                           
MHO~981 (345)&OF361A  & 20:17:40.5 & +39:20:32  & $10''$\\
       &OF361B  & 20:17:40.4 & +39:20:47  & $5''$\\
       &OF361C  & 20:17:40.4 & +39:20:59  & $5''$\\
\hline                           
MHO~982 (344)&OF360A  & 20:17:46.7 & +39:20:46 & $3''$\\
       &OF360B  & 20:17:44.9 & +39:20:33  & $3''$\\
\hline
\end{tabular}
\end{table*}

The most prominent detail OF373C as well as another feature, OF376A, coincide with diffuse emission in near-IR {\it Spitzer} bands, including $4.5\,\mu$m (Fig.~\ref{fig:EGO}, bottom), which supposedly contains emission from shocked molecular hydrogen \citep{2005ApJ...630L.181R,2005ApJ...634L.185B}. Diffuse $4.5\,\mu$m emission objects are known as extended green objects (EGOs) and are believed to be signposts of outflow activity. Their association with the H$_2$ emission features is shown in Fig.~\ref{fig:EGO} (bottom), which is a three-color representation of the protocluster area using a conventional color-coding based on $3.6\,\mu$m (red), $4.5\,\mu$m (green), and $8\,\mu$m (blue) {\it Spitzer} bands. It is seen that of all the suggested outflow features and associated MHOs only two, OF373C and OF376A, are clearly associated with EGOs.

\begin{figure}
\centering{\includegraphics[width=0.8\textwidth]{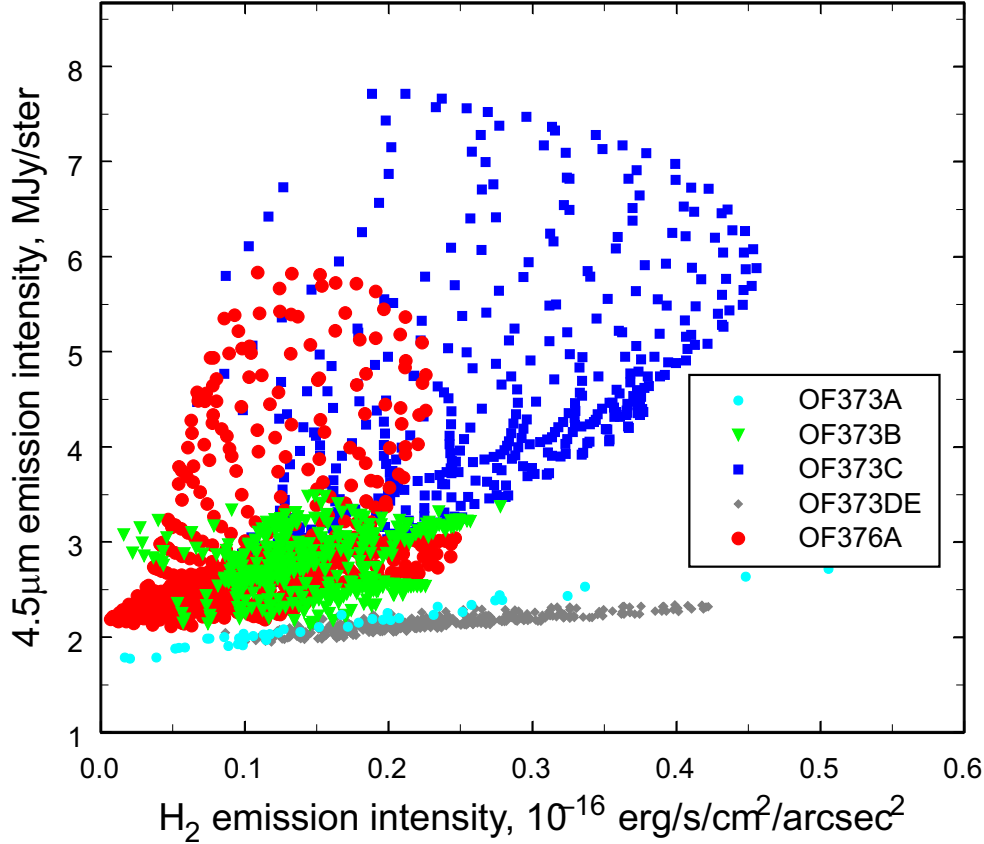}}
\caption{Relation between molecular hydrogen emission and 4.5\,$\mu$m emission in the selected regions of the protocluster area.}
\label{fig:pc45mum}
\end{figure}

Another way to demonstrate this is to consider a pixel-by-pixel relation between molecular hydrogen emission and 4.5\,$\mu$m emission in the selected MHOs, shown in Fig.~\ref{fig:pc45mum}. Features OF373A and OF373DE (as well as OF376B and OF374C, which are not shown to avoid cluttering) show 4.5\,$\mu$m emission at a background level, while being quite bright in 2.12\,$\mu$m molecular hydrogen emission. Emission at 4.5\,$\mu$m in OF373B feature (green filled triangles) slightly exceeds a background level, but also does not show any correlation with 2.12\,$\mu$m emission. Features OF373C (blue filled squares) and OF376A (red filled circles), on the contrary, show{ 4.5\,$\mu$m and 2.12\,$\mu$m intensities, which seem to be correlated, but with a significant scatter and a complicated shape of the point distribution on the plot. This complicated shape may indicate the presence of some small-scale irregularities in the distribution of 4.5\,$\mu$m and 2.12\,$\mu$m intensities, like displacements mentioned previously in relation to the infrared blobs. Still, we believe that the features OF373C and OF376A contain possible outflows from massive protostellar clumps}.

It is interesting that OF373A--OF373DE are designated as features of the same outflow by \cite{2018ApJS..234....8M}, but in Fig.~\ref{fig:pc45mum} they follow different trends, and the same is true for OF376A and OF376B as well. We also note that if OF373A--OF373DE features are indeed part of the same jet, at a distance of vdB~130 that would result in a projected end-to-end jet length well over 1 pc and place it among the longest young stellar object (YSO) jets seen in H$_2$ emission \citep{2006ApJ...636L.141H,2018ApJS..234....8M}.

While in the vicinity of the infrared blobs, discussed in the previous subsection, we were mostly concerned with extended H$_2$ emission, we also note a presence of compact H$_2$ emission sources between the shell and the blobs, indicated on Fig.~\ref{fig:H2emisshellblobs}. These sources are associated with two outflows, which are listed in the catalog by \citet{2018ApJS..234....8M} as OF360 and OF361.

\section{Sequential star formation in the vicinity of the vdB~130 cluster?}

\begin{figure*}
\centering{\includegraphics[width=0.8\textwidth]{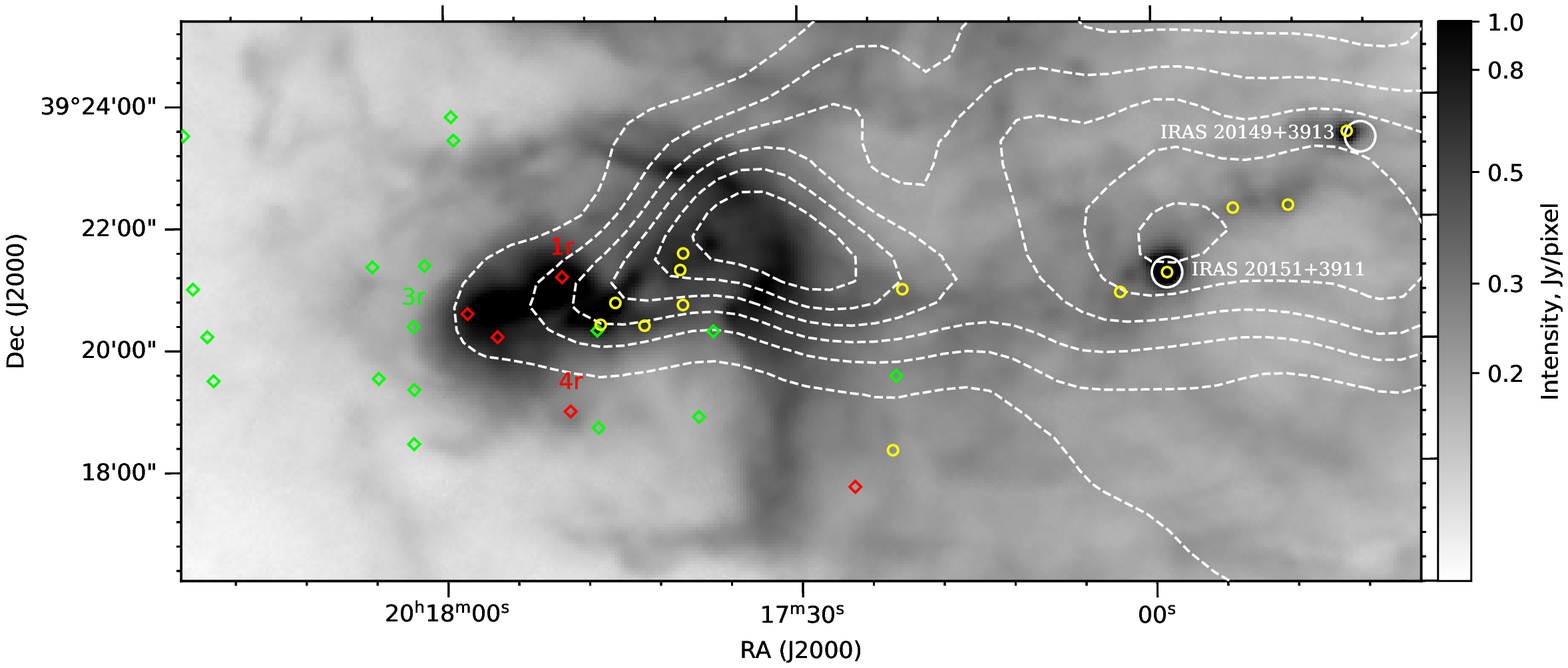}
\includegraphics[width=0.8\textwidth]{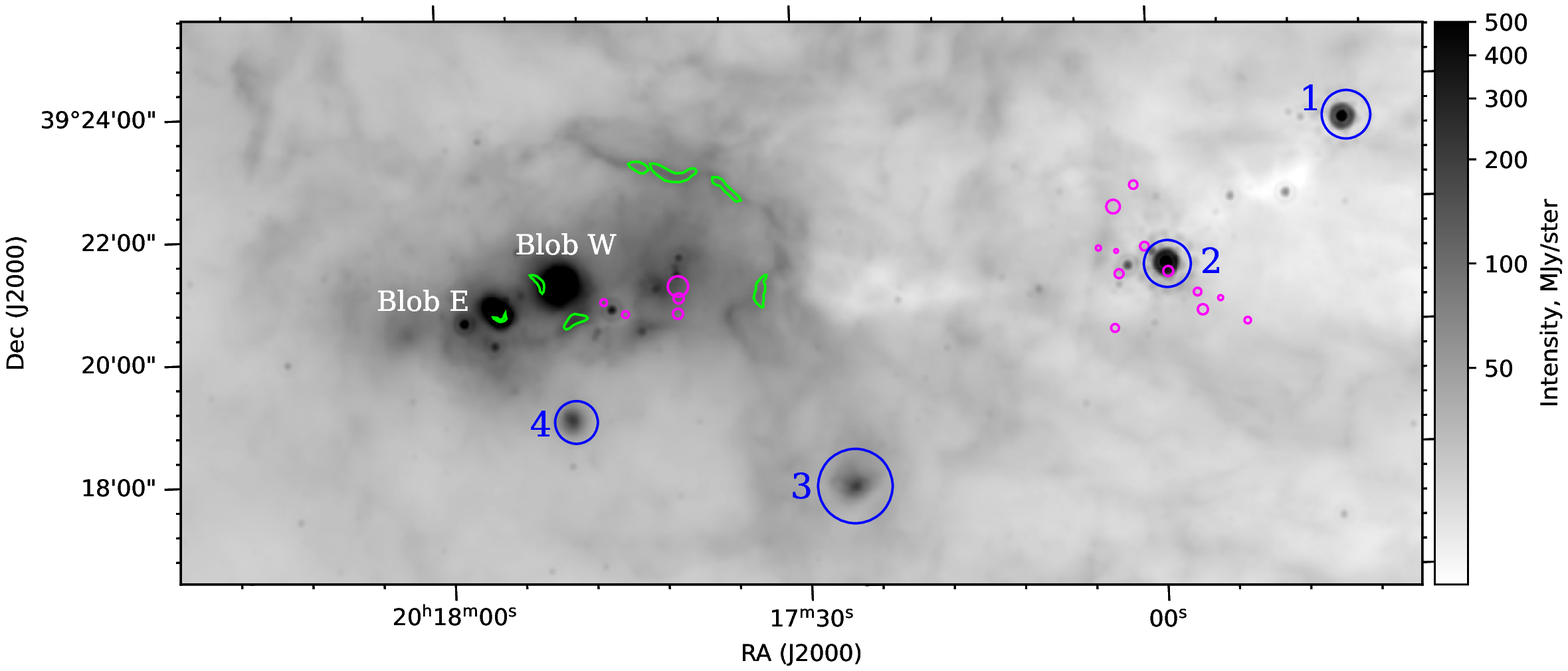}}
\caption{Variety of star-forming objects in the vicinity of vdB~130. Top: Background shows a map of $160\,\mu$m emission. YSOs identified in \citet{rastorguev_companion_paper} as vdB~130 cluster members are shown with red (Class I and Flat Spectrum) and green (Class II/III) diamonds. Yellow circles are YSOs found by \citet{2014AJ....148...11K}. White circles indicate ultra-compact \HII\ regions. Bottom: The same region in the 24\,$\mu$m {\it Spitzer} band. Green contours indicate regions of bright H$_2$ emission studied in this work. Blue circles show dense clumps associated with enhanced 24\,$\mu$m emission. Outflow features \citep{2018ApJS..234....8M} are shown with magenta circles.}
\label{outline}
\end{figure*}

The vdB~130 cluster and the surrounding area host a multitude of objects at various stages of prestellar and early stellar evolution \citep{rastorguev_companion_paper}. In $^{13}$CO emission, the entire region is dominated by a molecular cloud, having a length of about 12 pc, which is mostly located outside of the Cyg~OB1 supershell, but also protrudes into the ionized cavity as a molecular pillar having a cometary shape and facing east \citep[Cloud~A, ][]{sch07}. The outline of this cloud is shown in Fig.~\ref{outline} (top) both with CO emission contours and as a gray-scale map of 160\,$\mu$m emission, based on data obtained with the PACS instrument \citep{2010A&A...518L...2P} on-board of the {\it Herschel} Space Observatory \citep{2010A&A...518L...1P}. This pillar is directed toward nearby Cyg~OB1 stars, but this is most probably a mere coincidence as the pillar seems to be an eastern tip of a long molecular filament, visible in Herschel maps as an emission feature and in Spitzer IRAC and MIPS maps as an absorption feature (an infrared dark cloud; Fig.~\ref{outline}, bottom). The whole filament extends far beyond the supershell and, thus, could not have been shaped by ionizing radiation from Cyg~OB1 stars. It should be noted that at least some non-ionizing UV radiation leaks into the unperturbed region as revealed by diffuse $8\,\mu$m emission visible beyond the shell (Fig.~\ref{fig:cloud-proto}). While this radiation can alter the thermal state of the filament to a certain degree, causing its compression, it is hardly responsible for its formation and shaping, as such filaments are now believed to be a common feature of the molecular cloud evolution and the star formation process \citep{2022arXiv220503935P}.

The star formation might have been accelerated and orchestrated in the eastern part of the filament by its interaction with the expanding supershell. Specifically, the shock wave associated with the expanding shell might had compressed a pre-existing clump within the molecular filament, speeding up its collapse and the birth of the vdB~130 cluster. Hottest cluster stars have carved their own cavity, which probably resides at the observer's side of the filament as we see a shell encircling vdB~130, some emission, including diffuse H$_2$ emission, associated with the back wall, but no extinction features in front of the core of the vdB~130 cluster, which would indicate the presence of some foreground material. The filament itself and the associated molecular cloud have probably been partially disrupted by the infrared shell formation.

The eastern wall of the shell swept-up by the vdB~130 cluster stars might have compressed other dense clumps within the eastern part of the filament in its own turn, giving rise to Blob~E and Blob~W, which are currently still embedded in this side wall, which partially shows up as an extinction feature that engulfs Blob E and extends to the south of it (see Fig. 9b in \citetalias{sit15}). Our H$_2$ observations indicate that the infrared blobs are ultra-compact or maybe even hyper-compact \HII{} regions, so they are apparently younger than the half-opened bubble around the vdB~130 cluster.

Some other dense clumps are visible along the filament in its unperturbed part, to the west of the IR shell, which stand out as patches of enhanced IR ($24-160\,\mu$m) emission. These clumps host IRAS 20149+3913 and IRAS 20151+3911 sources identified as ultra-compact \HII\ regions by \citet{Bronfman}. The \HII{} regions are visible in the CS $(2-1)$ line at radial velocities $V_ {\rm LSR}$ equal to 3.1 and 4.2 km/sec typical of the cometary cloud (\citealt{sch07}, \citetalias{sit15} and \citetalias{sit19}). In Fig.~\ref{outline} (top), we also show protostars identified by \citet{2014AJ....148...11K} (yellow circles). Two of these protostars are apparently associated with the IRAS sources; the protostar associated with IRAS 20151+3911 probably drives the outflow OF373. Some other outflow driving sources, marked with magenta color in Fig.~\ref{outline} (top), also coincide with protostars from the work of \citet{2014AJ....148...11K}.

Red and green diamonds show, respectively, Class I and Flat Spectrum YSOs and Class II/III YSOs, probably, associated with vdB~130 as described in \citet{rastorguev_companion_paper}. They are mostly concentrated within the infrared shell with only a few of them being located outside. Our estimates show that the YSOs in the protocluster area are not gravitationally bound to vdb~130. They probably represent more scattered population of this star-forming region.

In the bottom panel of Fig.~\ref{outline} we show the same region mapped in the 24\,$\mu$m {\it Spitzer} band. On this panel we mark H$_2$ emission regions studied in this work along with four dense clumps associated with enhanced 24\,$\mu$m emission. As this emission is most probably generated by stochastically heated small grains absorbing UV photons, it also hints at the presence of young massive stars. Two of these regions, Clump~1 and Clump~2, located in the molecular cloud tail, coincide with regions of bright $160\,\mu$m emission (Fig.~\ref{outline}, top) and are also visible as enhancements on a column density map \citepalias[see Fig. 13 in ][]{sit19}. Clump 2, coincident with IRAS 20151+3911, is located in the direction of the compact protocluster core and of the peak in the CO emission. It is also associated with the possible EGOs OF373 and OF376 (denoted by green contours in Fig.~\ref{outline}, bottom). A part of the filament connecting the infrared shell, IRAS 20149+3913, and IRAS 20151+3911 is seen as a infrared dark cloud both on the 8\,$\mu$m and 24\,$\mu$m emission maps. There is no obvious protostar concentration toward Clump~1. To the south of vdB~130 there are two other infrared nebulosities, Clump~3 and Clump~4. Location of Clump~4 (which is also indicated in Fig.~\ref{fig:vdB130}, top) coincides with the cluster star~4r, so it may represent yet another small-scale star-forming site within the greater complex. Without the distance information the relation of Clump~3 to the studied region is unclear.

Two outflows with their associated MHOs along with several YSOs are located in dense parts of the filament in the interior of the infrared shell (in projection), but in general star formation seems to be at a more advanced stage in the vicinity of the vdB~130 cluster, which also shows up in the different properties of H$_2$ emission. In a close vicinity of vdB~130 we observe several extended diffuse H$_2$ emission features, which correlate with bright PAH emission features and are clearly associated with \HII\ regions, denoted as Blob~E, Blob~W, and H$\alpha$~Blob in our previous works. The shell around the vdB~130 cluster (H$\alpha$~Blob) appears to be quite developed and shows signs of being affected by the large-scale Cyg~OB1 irradiation in its H$_2$ line emission. Blob~E and Blob~W seem to be younger (despite of being closer to Cyg~OB1), and their H$_2$ emission is similar to that from a more compact \HII{} region. To the west of the IR shell, H$_2$ emission is only observed in several compact spots, with some of them being probably associated with jets. No signs of \HII{} regions similar to Blob~E and Blob~W are visible there.

In \citetalias{sit19} we proposed an evolutionary scenario, which relates the star formation activity in the vicinity of vdB~130 cluster to the expansion of the Cyg~OB1 supershell and its interaction with the cometary CO cloud, which represents a front end of the long molecular filamentary structure. According to this scenario, the cluster itself represent a most advanced star formation site, which lies in the center of the smaller scale infrared shell, encompassing the cluster from the west. To the east of the cluster, the expansion of the \HII\ region around the cluster might have stimulated star formation and subsequent \HII\ region formation in Blob W and Blob E (and, maybe, in Clump~4 around the 4r star). The infrared shell seemingly has not yet triggered any noticeable star formation activity (except for, maybe, Clump~3).

The general structure of H$_2$ emission is broadly consistent with this scenario, but few cautionary notes are in order. First, if the vdb~130 cluster age is indeed 10~Myr, its formation could not have been triggered by the expansion of the Cyg~OB1 supershell. A projected distance from the cluster to the supershell is of the order of at most a few parsec. This is much smaller than the distance that the supershell have traversed in 10~Myr. In this case, the supershell might have just cleaned up the vicinity of the cloud head from the less dense molecular gas and revealed the pre-existing star-forming cluster. However, as we argued above, 10~Myr can only be considered as an upper limit of the cluster age. The presence of numerous YSOs, including those of Class~I and Class~II, that are probable cluster members \citep{rastorguev_companion_paper}, indicates that the actual age of vdB~130 can be much smaller, being comparable to the time since the passage of the supershell.

Star formation in the quiescent (undisturbed) part of the filament, including the protocluster formation, is probably caused by some other events unrelated directly to the supershell expansion. One possibility is that the protocluster formation has been somehow triggered or accelerated by non-ionizing UV radiation of the association O-stars, for example, by compressing pre-existing dense clumps \citep{1999ARep...43..645S}. This radiation may indeed penetrate the supershell, which is emphasized by the presence of diffuse emission in near-IR {\it Spitzer} bands beyond the supershell (see Fig.~\ref{fig:cloud-proto}). Such an option may be explored further with observations of molecular lines \citep[e.g.][]{2008A&A...489..207S}.

Finally, it is possible that the region has undergone several episodes of star formation triggering, driven by mutual interactions of accretion flows, ionized regions, and (somewhat later) supernova explosions from the Cyg~OB1 stars. At earlier stages of the star-forming complex evolution this may indeed produce ``oscillating'' \HII{} regions that grow and shrink repeatedly \citep{2010ApJ...719..831P} leading to advance and retreat of the associated ionizing radiation. In such a case, the vdB~130 formation and the protocluster formation might have been caused by two independent triggering events. But at later evolutionary stages after the supershell formation such a complex behavior hardly can be expected.

Generally speaking, all the listed evidences do not contradict the scenario of the sequential star formation in the vdB~130 area. However, we must admit that they do not prove it with certainty. Much more detailed data on the region are needed to make more sound conclusions.

\section{Conclusions}

We present results of new observations of the vbB~130 cluster vicinity performed with a filter centered on the H$_2$ line at 2.12\,$\mu$m installed on the 2.5-metre telescope of the CMO MSU. It is shown that in the vicinity of the vdB~130 cluster molecular hydrogen ro-vibrational emission has low contrast but is clearly spatially coincident with 8\,$\mu$ emission related to PAH particles both in the infrared shell and in the regions around infrared blobs. There is a positive correlation between the H$_2$ line and 8\,$\mu$m band intensity almost everywhere except for the inner part of the brightest H$_2$ emission feature surrounding Blob~E. This indicates that H$_2$ emission in the vdB~130 cluster vicinity mostly represents fluorescent emission, excited by the UV radiation of the cluster stars in numerous PDRs, revealing complex gas distribution in the area. Compact patches of H$_2$ emission in the protocluster area are probably collisionally excited and trace several protostellar outflows. 

The observational results do not contradict the scenario of sequential star formation stimulated by an expanding supershell around the Cyg OB1 association, but on the other hand are not sound enough to claim it with certainty.

\begin{acknowledgments}
This work is partially based on the data from the Spitzer Space Telescope
operated by the Jet Propulsion Laboratory, California Institute of
Technology, under NASA contract 1407, and the data from the  2MASS catalogue
(University of Massachusetts, California Institute of Technology, NASA and
NSF). The Spitzer GLIMPSE360 images are shown courtesy of NASA (National Aeronautics and Space
Administration)/JPL-Caltech. CO observational data have been kindly provided by N.~Schneider \citep{sch07,2011A&A...529A...1S}.
The MHO catalog is hosted by the University of Kent.
{\it Herschel} is an ESA space observatory with science instruments provided by European-led Principal Investigator consortia and with important participation from NASA.
\end{acknowledgments}

\section*{FUNDING}

This study was carried out using the equipment bought with the funds of the Program of the Development of M.V. Lomonosov Moscow State University and supported by the RFBR grants (18-02-00976, 18-02-00890, 19-02-00611). D. Wiebe, A.M. Tatarnikov, and A.A. Tatarnikov were supported by the RFBR grant 20-02-00643. Authors acknowledge the support from the Program of Development of M.V. Lomonosov Moscow State University (Leading Scientific School ``Physics of stars, relativistic objects and galaxies''). The work of A. Topchieva was supported by the Foundation for the Advancement of Theoretical Physics and Mathematics ``BASIS''.

\section*{CONFLICT OF INTERESTS}

The authors declare the absence of the conflict of interests.

\bibliographystyle{aspb1}
\bibliography{wiebe}

\end{document}